# Lithography-Free, Manganese-Based Ultra-Broadband Absorption through Annealing-Based Deformation of Thin Layers into Metal-Air Composites

Majid Aalizadeh,[1,2,*] Amin Khavasi,[3] Bayram Butun,[2] Ekmel Ozbay[1,2,4,5]

[1] Department of Electrical and Electronics Engineering, Bilkent University, Ankara 06800, Turkey
[2] Nanotechnology Research Center (NANOTAM), Bilkent University, Ankara 06800, Turkey
[3] Electrical Engineering Department, Sharif University of Technology, Tehran, 11155-4363, Iran
[4] National Nanotechnology Research Center (UNAM), Bilkent University, Ankara 06800, Turkey
[5] Department of Physics, Bilkent University, Ankara 06800, Turkey
*majid.aalizadeh@bilkent.edu.tr

## Abstract

Fabrication, characterization, and analysis of an ultra-broadband lithography-free absorber is presented. An over 94% average absorption is experimentally achieved in the wavelength range of 450-1400 nm. This ultra-broadband absorption is obtained by a simple annealed tri-layer metal-insulator-metal (MIM) configuration. The metal used in the structure is Manganese (Mn), which also makes the structure cost-effective. It is shown that the structure retains its high absorption for TM polarization, up to 70 degrees, and, for TE polarization, up to 50 degrees. Moreover, the physical mechanism behind this broadband absorption is explained. Being both lithography-free and cost-effective, the structure is a perfect candidate for large-area and mass production purposes.

## 1. Introduction

Broadband absorbers (BBAs) have gained a lot of interest in the recent decade due to their vast range of applications in photovoltaics [1], photodetection [2], thermal emission [3], thermal imaging [4], shielding [5], etc. In the design and fabrication of BBAs, lithography is considered as a serious barrier on the way to their mass and high-throughput production. Therefore, it has always been very desirable to come up with lithography-free, cost-effective, and easy-to-fabricate structures. Some of the designs for lithography-free absorbers are based on planar MIM cavities [6,7], metal-insulator multilayer stacks [8], chemically synthesized metal-coated dielectric nanowires [9], random nanopillars [10], and etching-based random pyramids formed in doped Silicon wafers [11]. In addition, lithography-free absorbers without a broadband response are also of high interest, and they cover a wide range of crucial applications such as filtering [12], real-time tuning [13], and bio-sensing [14].

In this work, through annealing the MIM structure, we experimentally obtain over 94% average absorption in the ultra-wide range of 450-1400 nm that is a very promising result for a simple tri-layer lithography-free absorber. It is shown that annealing the sample leads to more than a twofold enhancement in the absorption bandwidth. We show that this is due to the fact that, after annealing, the top layer deforms into a metal-air composite with randomly sized, distributed and shaped nanoholes that support different Localized Surface Plasmon (LSP) resonance modes. The collective effect of the resonances of all random nanoholes leads to a significant absorption enhancement. The other feature contributing to such a broad absorption by such a simple structure is the optical properties of Mn that makes it an excellent candidate for broadband absorbers [6]. We experimentally demonstrate the absorption results for both cases of before and after annealing, and for a variety of incidence angles at both polarizations. This structure can serve as a very good candidate for large-area production of cost-effective absorbers covering a wide range of applications.

## 2. Calculation and Theory

Let us start with investigating the physical mechanism of the MIM resonant absorbers. Figure 1(a) demonstrates the schematic of the MIM configuration. The thickness of the layers from top to bottom is denoted as dt, di, and db. The top layer must be optically thin enough to allow an appropriate portion of the incident beam penetrate into the cavity, and eventually get coupled to the cavity resonance mode. The resonant wave of the cavity for which the Fabry-Perot resonance condition is satisfied, will bounce back and forth between the top and bottom metallic layers. In each reflection a portion of the resonating wave is absorbed, which eventually leads to the perfect absorption of the light trapped inside the cavity.

The Fabry-Perot resonance condition is as follows [15]:

$$2\left(\frac{2\pi}{\lambda_{res}}\right)n_i d_i + \varphi_b + \varphi_t = 2\pi m \tag{1}$$

where $\lambda_{res}$, $n_i$, $\varphi_b$, $\varphi_t$, m denote the resonance wavelength, refractive index of the dielectric material, phase-shift due to the reflection from the bottom metal, phase-shift due to the reflection from the top metal, and an integer number corresponding to the resonance order, respectively. The term $n_i d_i$ is the optical beam path traveled by the beam trapped inside the cavity, in each half-round of a full round-trip. It can be observed from the equation that nidi (or the dielectric layer) mainly determines the spectral location of the absorption band, i.e. $\lambda_{res}$. It is noteworthy that typically, dielectric oxides such as $SiO_2$, $Al_2O_3$, $HfO_2$, and $TiO_2$ have an almost constant refractive index in our desired wavelength range (visible and NIR). This means that for fixed metal layers, one can almost freely choose the dielectric material which determines $n_i$, and accordingly adjust the value of $d_i$ to obtain the desired beam path inside the cavity, and consequently determine the absorption band location. In other words, different dielectric oxides can lead to the same optical beam path and absorption band location, by having the needed thickness.

Since we can freely choose the dielectric oxide as explained above, we have chosen to use $Al_2O_3$. The main reason for this choice is that this oxide can be coated using the atomic layer deposition (ALD) method. ALD devices perform atomic layer-by-layer depositions that in comparison with the physical evaporation-based deposition methods, result in much more conformal thin films. Moreover, by using this method, we have control over each atomic layer of the depositing material which provides us with a very high precision in obtaining the desired thickness [16]. It is noteworthy that the material and the thickness of the bottom and top metal layers also determine the values of $\varphi_b$ and $\varphi_t$, respectively, thus affecting $\lambda_{res}$.

Now, in order to have a broadband absorption, the quality-factor of the resonance mode must decrease sufficiently. To achieve this objective, highly-lossy metals are required to be used in the MIM configuration. In our earlier work, it was experimentally demonstrated that Mn has a superior performance in the MIM broadband absorbers compared to the other appropriate candidates for broadband absorption including Cr, Pt, Ti, and W [6]. It was explained that this is due to the fact that Mn has a small real part of permittivity which leads to high field penetration and has a large imaginary part of permittivity which is the cause of strong absorption. Therefore, Mn is chosen for both metal layers in our MIM configuration. In addition, unlike expensive metals such as Platinum, Gold, or Silver, Mn is an economically cost-effective and environmentally-friendly metal which has the potential to be used for the large-scale and mass production of this device [17,18].

Simulations have confirmed that, to have strong absorption in the desired wavelength range that starts from visible and exceeds to NIR, for the case of using Mn and $Al_2O_3$, the optimal dimensions are as $dt = 5$ nm and $di = 70$ nm. Obviously, increasing the value of di leads to a red-shift in the absorption band (see Eq. 1), and decreasing it exerts a blue-shift (similar changes in $n_i$ also lead to similar results). However, in our previous work, it has been shown in detail that the absorption of the broadband MIM structure has a good fabrication tolerance for relatively small changes in the value of di. In other words, changes in the order of a few nanometers seem not to have any significant effect on the absorption band and the response of the structure is almost robust to minor deviations of the dielectric layer thickness from its desired value [6]. The bottom layer only needs to be sufficiently thick enough to block any transmission so that we have conservatively chosen $d_b = 200$ nm. Absorption (A) in the general case is calculated from A=1-R-T where R and T denote the reflection and transmission, respectively. In our case, T is zero so the absorption can be simply calculated from A=1-R.

## 3. Results and discussions

The measured and calculated absorption spectrum of the absorber before annealing, at normal incidence is shown in Fig. 1(b). The calculations are carried out using the circuit model, the details of which can be found in [6]. The refractive index values of $Al_2O_3$ and Mn used in the calculations are obtained by fitting the spectroscopic data from ellipsometry measurements. It can be observed that the calculations are in good agreement with the experimental results. The structure experimentally gives above 94% average absorption in the broad range of 450-900 nm. The scanning electron microscope (SEM) image of the cross-section of the structure milled by focusing ion beam (FIB) is shown in the inset of Fig. 1(b). Measured absorptions for TM and TE polarizations with oblique incidence are also shown in Figs. 1(c,d), respectively.

Moreover, Figs. 2(c,d) demonstrate the measured absorption spectrum for different angles of incidence and for TM and TE polarizations, respectively. It can be observed that the structure maintains its high absorption response for the TM polarization up to large incidence angles as high as 70 degrees. However, TE polarization seems to be more sensitive to the variations of the incidence angle. The oblique incidence reflection measurements along with the spectroscopic data extractions are carried out using J.A. Woollam Co. Inc. VASE Ellipsometer.

In order to obtain randomly sized and distributed nanoholes in the top metal layer that can support LSP resonances which lead to the absorption enhancement, we can use the dewetting method [1]. Dewetting is a lithography-free process in which the ultra-thin metallic layers get annealed at high temperatures to obtain nanopatterns (nanoholes or nanoislands, based on the annealing recipe) in them. The randomness of the size, shape, and distribution of the nanoholes leads to the existence of various LSP modes with different spectral positions, the superposition of which, contributes to the broadening of the absorption band.

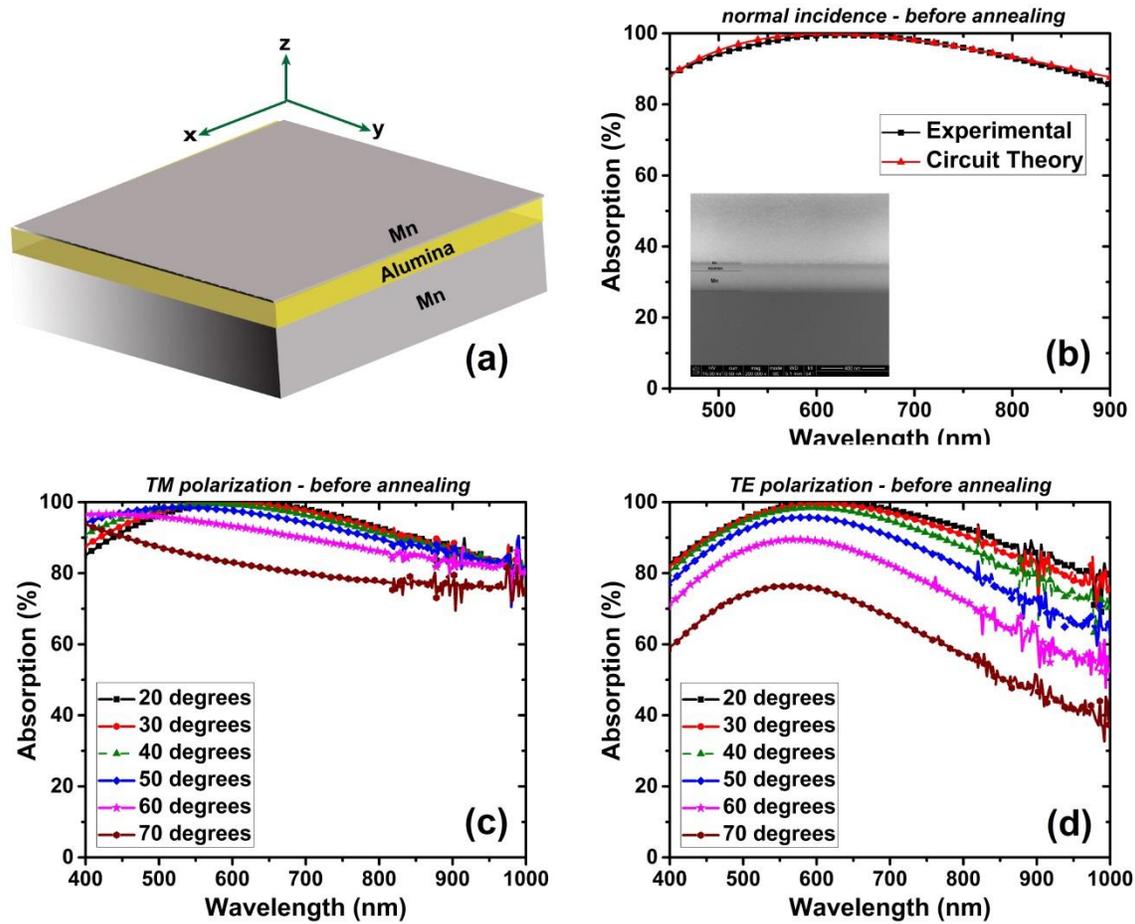

Fig. 1. (a) Schematic of the MIM absorber, (b) Measured and circuit theory-based calculated absorption of the structure for normal incidence, inset: SEM image of the cross-section of the fabricated MIM absorber, milled by FIB. Measured absorption spectra with different angles of incidence for (c) TM and (d) TE polarization. All of the figures correspond to the MIM sample before annealing.

To obtain nanoholes in the top thin Mn layer, we performed the dewetting process on our fabricated MIM samples by examining different annealing recipes. The best result, which is presented in this letter, belongs to the case of annealing the sample for 5 minutes at a temperature of 500℃. The rise time to reach that temperature was fixed to 5 minutes and the temperature falling time (sample cooling period) was set to 1 minute. It is noteworthy that annealing at lower temperatures resulted in an insignificant absorption enhancement, and higher temperatures led to losing the absorption strength compared to the best annealing recipe. The SEM image of the top view of the dewetted MIM sample is shown in Fig. 2(a). It can be clearly observed that there is a high randomness in the size, shape, and distribution of the nanoholes. The measured absorption spectrum of the annealed sample for the case of normal incidence is shown in Fig. 2(b). For the purpose of a clear visualization of the absorption band broadening, the measured absorption spectrum of the sample before annealing is also included in this figure. It can be seen that the absorption band has experienced a significant enhancement compared to before annealing. The absorption has above 94% average value in the broad wavelength range spanning from 450 to 1400 nm. Therefore, compared to before annealing, the upper limit of the absorption band of above 94% average absorption, is increased from 900 to 1400 nm, while the lower wavelength limit is fixed at 450 nm, i.e. a 500 nm or over twofold bandwidth enhancement is experimentally achieved. Such an achievement is obtained by only adding a simple and high-throughput annealing stage to the fabrication process. The normal incidence reflection measurements are carried out using a reflectometer and fiber optical cable-based home-made setup for up to 900 nm wavelength, and for larger wavelengths, measurements are done by a Fourier transform infrared (FTIR, Bruker) device. Inset of Fig. 2(b) shows the image of the annealed fabricated sample which appears with black color due to the strong light absorption.

Figures 2 (c,d) demonstrate the measured absorption for oblique incidence when the incident wave is TM and TE polarized, respectively. It can be observed that the absorption band for oblique incidence has also experienced a significant enhancement by annealing. The results show that, for the TM polarization, the absorption remains high for large angles up to 70 degrees, while for the TE polarization, the structure almost loses its high absorption performance after 50 degrees. In other words, the overall behavior of the structure for oblique incidence indicates that the structure has very low incidence angle sensitivity, especially for the TM polarization.

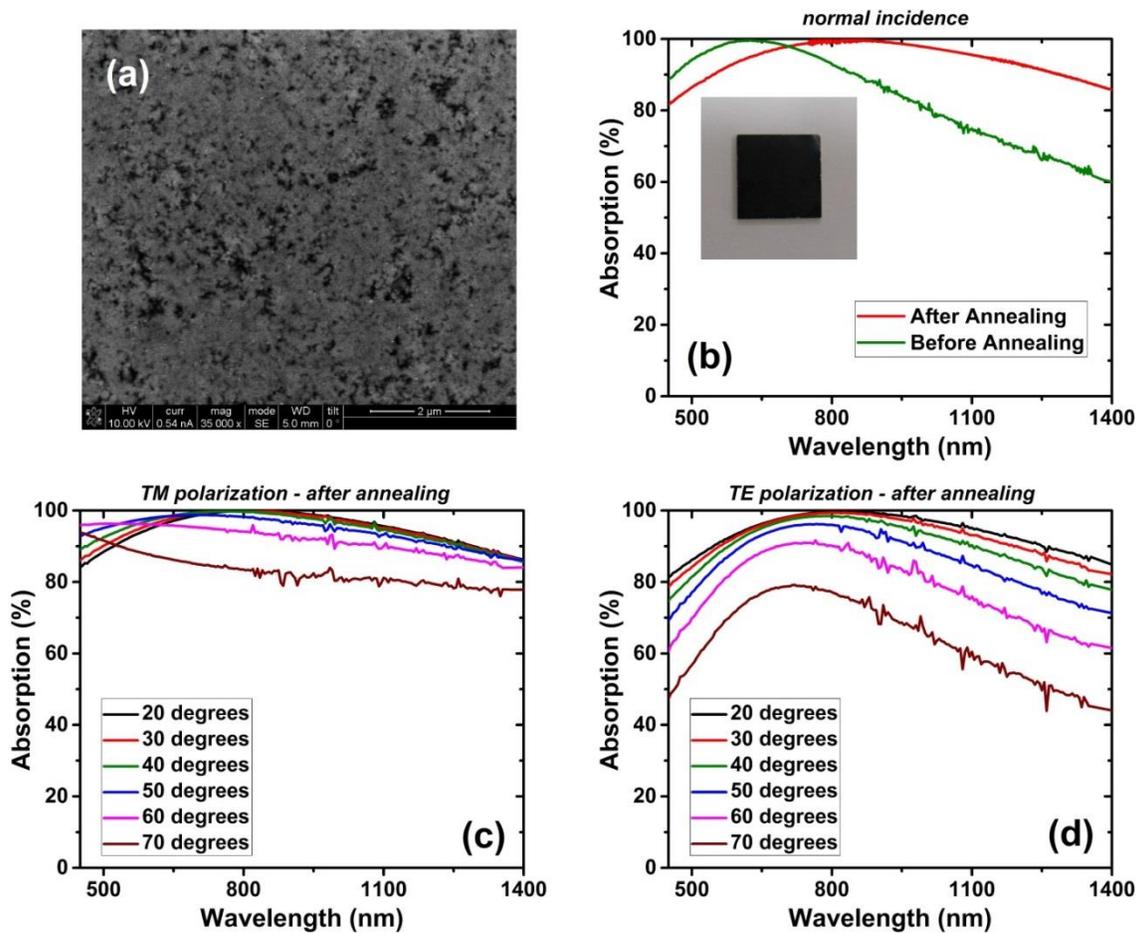

Fig. 2. (a) SEM image taken from the surface of the annealed MIM sample, (b) measured absorption of the sample before (green line) and after (red line) annealing for the case of normal incidence, and, for different angles of incidence in the cases of (c) TM and (d) TE polarizations. Inset of (b) shows the image of the annealed fabricated sample.

To observe the physical mechanism behind this broadband absorption, the SEM image of the annealed structure is imported into the Lumerical FDTD solver software [20] and the simulation is performed by defining it as the top layer. A square area of 1 μm2 from the SEM image is chosen for the simulation, and the periodic boundary conditions are defined in the both x and y directions. Figures 3(a,b,c) demonstrate the magnitude of the magnetic field, electric field, and absorption intensity, respectively. The field monitor is placed in the xy-plane cross-section at center of the top layer, with regards to the z-axis, and simulation results are shown for the four different wavelengths of 600, 800, 1000, and 1200 nm.

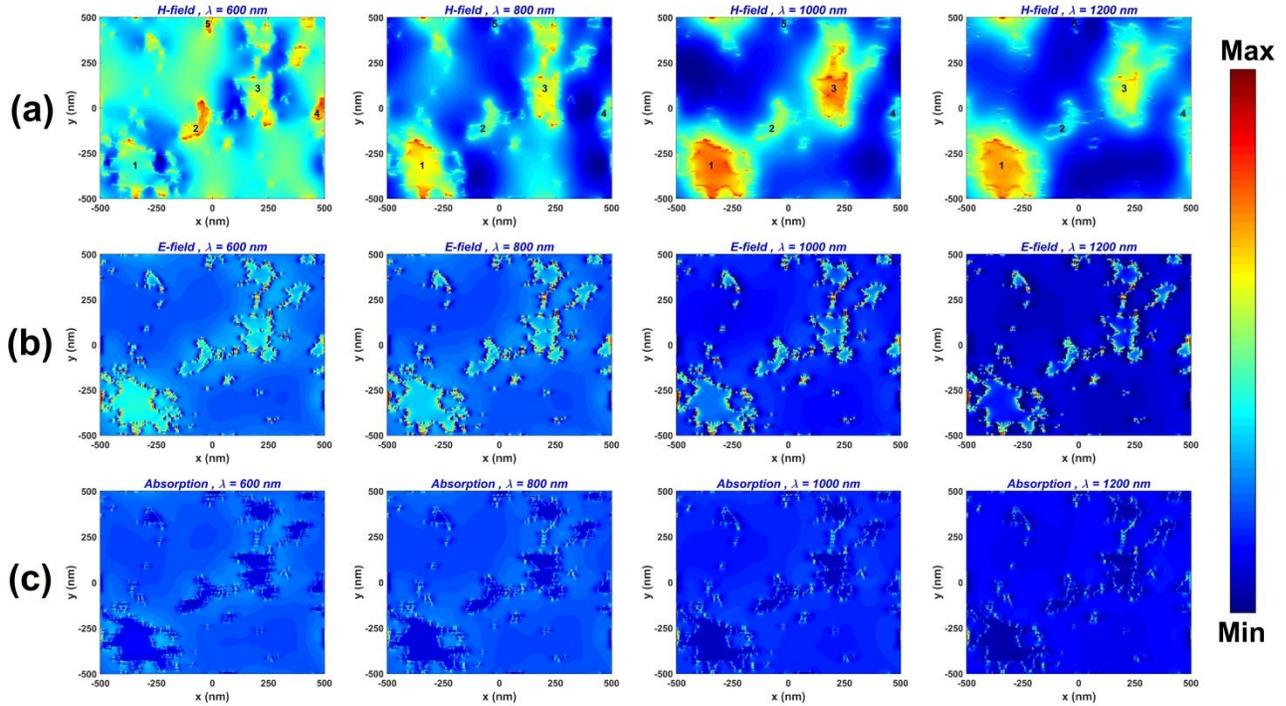

Fig. 3. (a) Magnetic field magnitude, (b) electric field magnitude, and (c) absorption intensity distribution of on the xy cross-section of the center of the top metal layer of the annealed structure, simulated by importing the SEM image of the sample into the Lumerical FDTD solver.

Magnetic field patterns shown in Fig. 3(a) provide a clear explanation for the role of random nanoholes in absorption band enhancement. As can be seen, the field gets localized, enhanced, and confined inside the nanoholes, which is demonstrative of LSP mode excitations. It can be observed that for the cases of different wavelengths, different nanoholes with different sizes get excited. To explain this phenomenon, relatively large and observable nanoholes are numbered from 1 to 5 in Fig. 3(a). It can be seen that for instance, at the wavelength of 600 nm, mainly nanoholes 2, 4, and 5 get excited and field localization at the other nanoholes is not as strong as these ones. On the other hand, for the wavelength of 1000 nm, mainly nanoholes 1 and 3 get excited. Therefore, as mentioned above, at any chosen wavelength in the absorption band, specific nanoholes with specific sizes get excited, and as the wavelength increases, it is observable that larger sized nanoholes get excited, which is in total agreement with the nature of LSP resonance modes [21]. Consequently, the superposition and collective effects of all randomly-sized nanoholes existing in the top metallic layer lead to a significant absorption band widening in a broad wavelength range.

It can be observed from the electric field magnitude pattern in Fig. 3(b) that there is a field localization and enhancement in the nanohole edges for all wavelengths. As will be clearly explained later, such a field localization at the nanoholes edges which include metal-air interfaces, leads to an absorption enhancement at these points.

The absorption intensity patterns shown in Fig. 3(c) are calculated using the following formula:

$$P_{abs} = -0.5\omega |E|^2 imag(\varepsilon) \tag{2}$$

where $\omega$ is the angular frequency, |E| is the electric field magnitude, and $\varepsilon$ is the permittivity of the material. It is obvious that the absorption will be zero for the air regions inside the nanoholes, and for a constant wavelength, i.e. constant $\omega$, the absorption intensity in the metallic parts with the same metal will only be a function of $|E|^2$. This is the reason why the absorption intensity patterns of the metallic parts in Fig. 3(c) are relatively similar to the electric field magnitude patterns in Fig. 3(b), except with the explained differences. It can be clearly observed that a significant absorption enhancement happens at the hotspots of the edges of the nanoholes on the top layer which is the result of field localizations due to the excitation of LSPs, shown in Fig. 3(a).

# 4. Conclusion

In summary, an ultra-broadband tri-layer MIM absorber is fabricated and characterized. We have shown that, through annealing, an above 94% average absorption is experimentally achieved in the wide range of 450-1400 nm, which is a very promising result compared to the simplicity of the structure. It is explained that this is due to the deformation of the top layer into metal-air composite that includes random nanoholes with the potential of supporting different of LSP resonance. Moreover, it is shown that at different wavelengths, different nanoholes get excited and a collective effect of all nanoholes contributes to such a significant absorption enhancement. It is experimentally demonstrated that this absorber has very low angle sensitivity for the TM polarization up to 70 degrees, and for the TE polarization up to 50 degrees. The structure is lithography-free and cost-effective, so it is a perfect candidate for large-area production and utilization in photovoltaic and solar cell applications.

## Funding

Turkish Academy of Sciences (TUBA); Research Office of Sharif University of Technology.

## Acknowledgments

Ekmel Ozbay acknowledges partial support from the TUBA. Amin Khavasi also acknowledges Research Office of Sharif University of Technology.